\documentclass[12pt]{article}
\usepackage{epsfig,amsfonts,amssymb}
\usepackage{cite}
\input epsf.sty
\usepackage{cite}

\def\ZZZ{{\hbox{ Z\kern-1.6mm Z}}}
\def\RRR{{\hbox{ R\kern-2.4mm R}}}
\def\CCC{{\hbox{ C\kern-2.0mm C}}}
\def\zzz{{\hbox{z\kern-1mm z}}}

\newcommand{\qeq}{{\hbox{=\kern-2.3mm ? \kern.5mm }}}
\renewcommand{\qeq}{=}

\newcommand{\wt}{\widetilde}
\newcommand{\wh}{\widehat}

\newcommand{\bd}{\bar{\rm D}}

\newcommand{\be}{\begin{equation}}
\newcommand{\ee}{\end{equation}}
\newcommand{\ben}{\begin{eqnarray}\displaystyle}
\newcommand{\een}{\end{eqnarray}}

\newcommand{\bea}[1]{\begin{eqnarray}\label{#1} }
\newcommand{\eea}{\end{eqnarray}}

\newcommand{\refb}[1]{(\ref{#1})}
\newcommand{\p}{\partial}
\newcommand{\sectiono}[1]{\section{#1}\setcounter{equation}{0}}

\def\one{{\hbox{ 1\kern-.8mm l}}}
\def\zero{{\hbox{ 0\kern-1.5mm 0}}}

\begin{document}

{}~
{}~

\vskip .6cm

{\baselineskip20pt
\begin{center}
{\Large \bf Tachyon Condensation on Separated 
Brane-Antibrane System} 

\end{center} }

\vskip .6cm
\medskip

\vspace*{4.0ex}

\centerline{\large \rm
 Arjun Bagchi and Ashoke Sen}

\vspace*{4.0ex}

\centerline{\large \it Harish-Chandra Research Institute}

\centerline{\large \it  Chhatnag Road, Jhusi,
Allahabad 211019, INDIA}

\vspace*{1.0ex}

\centerline{E-mail: arjun@hri.res.in, sen@hri.res.in}

\vspace*{5.0ex}

\centerline{\bf Abstract} \bigskip

We study the effect of tachyon condensation on a brane antibrane
pair in superstring theory
separated in the transverse direction. The static properties of
the tachyon potential analyzed using level truncated string field theory
reproduces the desired property that the dependence of the minimum
value of the potential on the initial distance of separation between
the branes decreases as we include higher level terms. The rolling
tachyon solution constructed using the conformal field theory
methods shows that if the initial separation between the branes is
less than a critical distance then the solution is described by an
exactly marginal deformation of the original conformal field
theory where the correlation functions of the deformed theory are
determined completely in terms of the correlation functions of the
undeformed theory without any need to regularize the theory.
Using this we give an expression for the pressure on the brane-antibrane
system as a power series expansion in $\exp(C x^0)$ for an
appropriate constant $C$.

\vspace*{4.0ex}

\vfill \eject

\baselineskip=18pt

\tableofcontents

\sectiono {Introduction} \label{s1}

The spectrum of the bosonic open string theory
 living on a D-brane is known to
have a tachyonic mode. 
We now have a good understanding of the physics around
the minimum of the tachyon potential, 
both via conformal field theory (CFT) methods\cite{9902105}, 
and numerical and 
analytical methods in string field 
theory\cite{9911116,9912249,0002237,0211012,
0511286, 0603159,0603195,0605254,0606131,0606142, 0610298,
0611110,0611200,0612050,0709.2888,0710.5358}.
In particular
it is  known that the tachyon potential
has a non-trivial minimum where the 
energy density from the potential 
exactly equals the negative of the D-brane tension
and as a result the 
sum vanishes. The
minimum represents a vacuum without any D-branes.
Using conformal field theory methods one can also study time
dependent solutions in string theory describing
the rolling of the tachyon towards
the vacuum\cite{0203211}.

Similar conjectures hold in the case of the superstrings 
where tachyonic modes
appear in unstable systems like non-BPS 
D-branes or brane-antibrane pairs\cite{9805170}. 
Level truncation gives 
numerical evidence for these conjectures in 
Berkovits superstring field 
theory\cite{0001084,0002211,0003220,0004015}, but
as of now we do not have 
an analytic solution for the vacuum.\footnote{An analytic
solution has recently been constructed 
in the superstring field theory based on the
cubic action\cite{0707.4591}.} As in the case of
bosonic string theory, one can also construct a conformal field
theory describing the rolling of the tachyon
towards the vacuum\cite{0203265}.

Most of the work on tachyon condensation in superstring field
theory has been carried out on an unstable D-brane system,
or a closely related system containing 
a coincident brane-antibrane pair.
In this paper, we look at a system of brane-antibrane 
pair separated
by a distance $d$. This is the configuration we expect to get
in any realistic situation 
involving tachyon condensation on a brane-antibrane system,
{\it e.g.}
in cosmology, where
the
brane-antibrane pair would start out separated 
from each other and gradually
come together by gravitational attraction\cite{9812483}. 
As they come closer than the critical distance the lowest lying
mode of the open string stretched between the brane and the antibrane
will become tachyonic and the condensation process would start. 
Thus if we want to study the end point of tachyon condensation for
such a system we need to study tachyon condensation on a separated
brane-antibrane pair.

Our analysis will be divided into two parts.
We first look at the static configuration of a 
separated brane-antibrane
pair, and carry out a level truncation analysis of the tachyon
vacuum using Berkovits' superstring field 
theory\cite{9503099,9912121}.
In this case we do not expect any surprise; rather we expect that at
the bottom of the potential the total energy 
density should continue to vanish
irrespective of the initial distance between the brane-antibrane
pair. This result is bourn out by our analysis. In particular
we find that while at the lowest level the value of the potential
at the minimum depends on the initial separation between the
brane-antibrane pair, this dependence reduces after inclusion of
higher level terms in the action.

The second part of the analysis involves study of the rolling tachyon
solution using conformal field theory method. 
Unlike in the case of rolling tachyon
on a non-BPS D-brane or a coincident brane-antibrane pair, 
in this case
we cannot construct an exact boundary state corresponding to
the time dependent configuration. Nevertheless using a perturbative
approach one can write down an expression for the pressure as a
series expansion in powers of 
$\exp(C x^0)$ for an appropriate constant $C$ depending on the
initial separation of the brane-antibrane system. We find that if
the initial separation between the brane-antibrane pair is less
than a critical distance then the coefficients of the various terms
of the expansion can be expressed in terms of non-singular integrals.
We analyze the behaviour of this series by computing the first few
terms in the expansion numerically.

For rolling tachyon solution
on a coincident brane-antibrane pair 
the final state was found to have
vanishing pressure but non-zero energy density\cite{0203265}. 
This reflects that
the final state is made of non-relativistic heavy closed string 
states\cite{0303139,0410103}.
If instead of starting with a coincident brane-antibrane pair we
begin with a separated brane-antibrane pair then the final state
in principle could be different, (say) consisting of a mixture
of non-relativistic heavy closed string states and 
radiation containing
relativistic light closed string states. Thus computation of
the final state
pressure is an important problem since this could tell us indirectly
about the composition of the final state.
Unfortunately since we only have a power series expansion for the
pressure, we cannot reach a definite conclusion about the final state
pressure. 
However we 
use the Pade approximant method to represent the known results
on the power
series expansion as a ratio of polynomial functions, and extrapolate
the result based on the first few coefficients to study
the behaviour of the pressure at large time. This naive
extrapolation gives results consistent with vanishing pressure at
late time.

During our analysis we also develop a general procedure for
studying rolling tachyon solution in situations where the tachyon
vertex operator is a non-trivial matter primary operator. We find that
as long as the tachyon is sufficiently tachyonic, \i.e.\ the tachyon
mass$^2$ is below a critical value, the system admits an exactly
marginal deformation describing the  rolling of the tachyon
away from the maximum. The essential point is that the integrated
vertex
operator describing a rolling tachyon deformation, obtained by
multiplying the zero momentum tachyon vertex operator by
$e^{C X^0}$ for an appropriate constant $C$, has non-singular
operator product with itself for sufficiently large $C$. As a result
deformation by this operator describes an exactly marginal deformation
of the conformal field theory.

\sectiono {Superstring field theory on brane-antibrane system}
\label{s2}

In this section we
give a quick review of the construction of the superstring
 field theory (SSFT)
on a 
brane antibrane pair. We then identify the specific components of the
string field which we shall use for the study of tachyon condensation on
a separated brane-antibrane pair.

 \subsection   {SSFT on a BPS D-brane}
We begin by looking at the GSO(+) sector of the superstring 
field theory which describes
the dynamics of the NS sector of open strings living on a 
single BPS D-brane. The CFT describing the first quantized
open string theory is a direct product of 
superconformal matter with $c=15$ containing the fields
$X^\mu$, $\psi^\mu$ for $0\le \mu\le 9$, 
and $b$, $c$, $\beta$, $\gamma$ 
ghost CFT with $c=-15$. The
$\beta$, $\gamma$ system  can be
reexpressed in terms of the bosonised ghosts $\xi$, 
$\eta$ and $\phi$ with\cite{FMS}
\begin{equation}
\beta = {\partial \xi}e^{-\phi} , \quad
\gamma = \eta e^{\phi}\, .
\end{equation} 
We shall be working in the large Hilbert space 
which includes the zero mode of the field $\xi$ and
use the convention set in \cite{0002211}. 
We normalize  the various fields so that the leading singularities
in the various operator product expansions have the following
form
\ben \label{erule} 
& \p X^\mu(z) \p X^\nu(w) \simeq - {1\over 2}
\eta^{\mu\nu}(z-w)^{-2}
\nonumber \\
& \p\phi(z) \p\phi(w) \simeq -(z-w)^{-2} \nonumber \\
& \psi^\mu(z) \psi^\nu(w) \simeq \eta^{\mu\nu}\,(z-w)^{-1}
\nonumber \\
&\xi(z) \eta(w) \simeq (z-w)^{-1} \nonumber \\
& b(z) c(w) \simeq (z-w)^{-1} \nonumber \\
&e^{\alpha\phi(z)} e^{\beta\phi(w)} \simeq (z-w)^{-\alpha\beta}
e^{(\alpha+\beta) w}, \quad z,w\in\CCC \nonumber \\
& e^{i k_1\cdot X(s)} e^{i k_2\cdot X(s')} \simeq
|s-s'|^{-2k_1\cdot k_2} e^{i(k_1+k_2)\cdot X(s')}, 
\quad s,s'\in\RRR\, .
\een
We shall denote by
$ \langle\prod_{i}A_{i}\rangle $ the
correlation functions in the combined 
matter-ghost boundary CFT (BCFT) 
on the unit disk with vertex operators $A_i$ inserted on
the boundary. 
The correlation functions are normalized as
\begin{equation}
<\xi (z) c \partial{c} \partial^{2}c (w) e^{-2\phi(y)}> = 2 \, .
\end{equation}    
The BRST operator is given by:
\begin{equation}
Q_{B} = \oint dz j_{B}(z) = \oint dz \lbrace 
c (T_{m} + T_{\xi\eta} + T_{\phi} ) +c \partial{c} b 
+ \eta e^{\phi} G_{m} - 
\eta \partial{\eta}e^{2\phi} b\rbrace
\end{equation} 
where
the $T$'s denote the energy momentum tensors 
for the various fields and $G_{m}$ is the matter 
superconformal generator:
\begin{eqnarray}
T_{m} &=& - (\partial X^{\mu}\partial 
X_{\mu} + \frac{1}{2} \psi^{\mu}\partial \psi_{\mu}) \nonumber\\
G_m &=& -i\sqrt 2\, \psi_\mu \p X^\mu \nonumber\\
T_{\xi\eta}&=& \partial {\xi} {\eta} \nonumber\\
T_{\phi}&=& - \frac{1}{2} \partial\phi\partial\phi -\p^2\phi\, .
\end{eqnarray} 

The Berkovits' superstring field theory action is given by
\begin{equation} \label{eact1}
S= \frac{1}{2g^{2}} \left\langle\left\langle 
(e^{-\Phi}Q_{B}e^{\Phi})\left(e^{-\Phi}{\eta}_{0}e^{\Phi}\right) - 
\int _{0}^{1} dt e^{-t\Phi} \partial_{t} e^{t\Phi}
\left\lbrace e^{-t\Phi}Q_{B}
e^{t\Phi}, e^{-t\Phi}{\eta}_{0}e^{t\Phi}\right\rbrace 
\right\rangle \right\rangle
\end{equation} 
where the string field $\Phi$ is a ghost number zero and
picture number 0 state of the CFT in the large Hilbert
space and the action needs to be defined by expanding 
\refb{eact1} in a power series in $\Phi$ and carefully 
preserving the order of the operators. 
The notation $\langle\langle
~\rangle\rangle$ means
\begin{equation}
\langle \langle A_{1} A_{2} ... A_{n}\rangle\rangle = 
\langle f_{1}^{(n)}\circ A_{1}(0) f_{2}^{(n)}\circ 
A_{2}(0) ... f_{n}^{(n)}\circ A_{n}(0) \rangle
\end{equation} 
with $ f_{l}^{(n)}
\circ A_{l} $ implying the 
conformal transformation of the operator 
$ A_{l} $ under the map $f_{l}^{(n)}$. 
The maps $f_{l}^{(n)}$ are given by
\begin{equation}
f_{l}^{(n)}(z) = e^{(2\pi i (l-1))/n} \left(\frac{1+ iz}{1- iz}
\right)^{2/n}
\end{equation} 

\subsection {SSFT on a non-BPS D-brane}

In order to extend this formalism to 
non-BPS D-branes one needs to take into 
account the GSO($-$) sector
that now comes into the picture. 
In order to keep the basic algebraic framework 
unchanged, one introduces
internal Chan-Paton (CP) factors and performs a 
trace over them. 
The GSO(+) sector states carry CP factor proportional to the
$2\times 2$ identity matrix $I$ whereas the GSO($-$) sector
states carry CP factor proportional to the Pauli matrix
$\sigma_1$.
Consequently, the complete string field $\wh\Phi$
is now represented by
\begin{equation}
\wh\Phi = \Phi_{+}\otimes I  + \Phi_{-}\otimes \sigma_{1}\, .
\end{equation} 
We also need to modify the $Q_B$ and $\eta_0$ 
operators by tensoring them with CP factor $\sigma_{3}$
\begin{equation}
\wh Q_B = Q_B \otimes \sigma_{3} , \hspace{2mm}    
\wh \eta_0 = \eta_0 \otimes \sigma_{3} \, .
\end{equation} 
In computing the double bracket
$\langle\langle~\rangle\rangle$ of hatted operators
we need to take the trace over the internal 
CP factors:
\begin{equation}
\langle \langle \wh A_{1} \wh A_{2} ... \wh A_{n}
\rangle\rangle = Tr \langle f_{1}^{(n)}\circ 
\wh A_{1}(0) f_{2}^{(n)}\circ \wh A_{2}(0) ... f_{n}^{(n)}
\circ \wh A_{n}(0) \rangle\, .
\end{equation}
The Berkovits' action looks 
almost the same, with all the fields and
operators replaced by hatted fields and operators respectively.
We also divide the action by an extra factor of 
2 in order to compensate for the factor of
2 coming from the trace over the internal CP factors:
\begin{equation}
S= \frac{1}{4g^{2}} \left\langle
\left\langle \left((e^{-\wh\Phi}Q_{B}e^{\wh\Phi})
(e^{-\wh\Phi}{\eta}_{0}e^{\wh\Phi}) - 
\int _{0}^{1} dt e^{-t\wh\Phi} \partial_{t}
 e^{t\wh\Phi}\lbrace e^{-t\wh\Phi}Q_{B}
 e^{t\wh\Phi}, e^{-t\wh\Phi}{\eta}_{0}
 e^{t\wh\Phi}\rbrace \right)\right\rangle \right\rangle
\end{equation} 

\subsection {SSFT on a D-brane-$\bar{\rm D}$-brane pair}

The formalism needs to be further modified in order to extend
it to the brane-antibrane system. Here besides the
internal CP factors like the ones used for the non-BPS 
branes one also needs to use external CP factors. There are four 
kinds of strings represented 
by the external CP matrices:
 \[A:\left(
	\begin{array}{cc}
	1 & 0 \\
	0 & 0
	\end{array}
\right) \> , \quad \> B:\left(
	\begin{array}{cc}
	0 & 0 \\
	0 & 1
	\end{array}
\right) \> , \quad \> C:\left(
	\begin{array}{cc}
	0 & 1 \\
	0 & 0
	\end{array}
\right) \> , \quad \> D:\left(
	\begin{array}{cc}
	0 & 0 \\
	1 & 0
	\end{array}
\right). \]
The strings on the individual branes are represented 
by the CP factors
$A$, $B$ or equivalently by $I$ and $\sigma_3$. 
These are in the GSO(+) sector. The GSO($-$) states 
are the ones which live on the strings stretched between 
the brane and the antibrane, -- 
they are represented by the CP factors $C$, $D$ or 
equivalently by $\sigma_1$, $\sigma_2$.
The complete string field $\wh\Phi$ now reads
\begin{equation}
\wh\Phi = \left( \Phi_{+}^{(1)}\otimes I + 
\Phi_{+}^{(2)}\otimes \sigma_{3}\right)\otimes I + 
\left( \Phi_{-}^{(3)}\otimes \sigma_{1} + \Phi_{-}^{(4)}
\otimes \sigma_{2}\right)\otimes \sigma_{1}\, .
\end{equation} 
We follow the convention that the external 
CP factor will be written first
followed by the internal CP factor.

The $\wh Q_B$ and $\wh\eta_0$ operators are now given by
\begin{equation}
\wh Q_B = Q_B \otimes I \otimes \sigma_{3} , \hspace{2mm}   
 \wh \eta_0 = \eta_0 \otimes I \otimes \sigma_{3} \, .
\end{equation} 
The double brackets
$\langle\langle~\rangle\rangle$
are now defined with a double trace, 
over both internal and external CP factors:
\begin{equation}
\langle \langle \wh A_{1} \wh A_{2} \ldots 
\wh A_{n}\rangle\rangle = Tr_{ext}\otimes Tr_{int} 
\langle f_{1}^{(n)}\circ \wh A_{1}(0) f_{2}^{(n)}
\circ \wh A_{2}(0) ... f_{n}^{(n)}\circ \wh A_{n}(0) \rangle\, .
\end{equation}
The action looks very much the same as for the non-BPS
D-brane except that we divide by a further
factor of 2 to compensate for the trace over the external Chan-Paton
factors:
\begin{equation}
S= \frac{1}{8g^{2}} \left\langle\left\langle \left((e^{-\wh\Phi}
Q_{B}e^{\wh\Phi})(e^{-\wh\Phi}{\eta}_{0}e^{\wh\Phi}) - 
\int _{0}^{1} dt e^{-t\wh\Phi} \partial_{t} 
e^{t\wh\Phi}\lbrace e^{-t\wh\Phi}
Q_{B}e^{t\wh\Phi}, e^{-t\wh\Phi}{\eta}_{0}
e^{t\wh\Phi}\rbrace 
\right)\right\rangle \right\rangle\, .
\end{equation} 
We shall find it convenient to consider 
the time direction as a circle with unit period. The 
tachyon potential would then just be the 
negative of the action for static configurations. 
With this normalization the total tension of the brane-antibrane
pair is given by
\be \label{etension}
T = {1\over 2\pi^2 g^2}\, .
\ee
For explicit calculations, it is useful to expand the action 
in a formal power series in $\widehat{\Phi}$. 
It can be arranged in the form\cite{0002211}
\begin{eqnarray} \label{expanded}
S =\frac{1}{4g^2}\sum_{M,N=0}^{\infty}\frac{(-1)^N}
{(M+N+2)!}
\left({M+N \atop N}\right)
\left\langle\left\langle\left(\widehat{Q}_B\widehat{\Phi}\right)
\widehat{\Phi}^M\left(\widehat{\eta}_0
\widehat{\Phi}\right)\widehat{\Phi}^N \right\rangle\right\rangle
\, .
\end{eqnarray}

\subsection {Separated D-branes}

 Our interest is in a configuration where the brane and the antibrane
 are separated from each other. In this case
 the mass of any state of the string
 stretched between the brane and the antibrane gets an 
 additional contribution from the tension of the string compared
 to the string whose both ends are on the same brane. If we denote
 by $d'$ the separation between the brane and the antibrane then
 in the $\alpha'=1$ unit this additional contribution, affecting
 the formula for the mass$^2$ in sectors C and D, is given by
 $d^{\prime 2}/4\pi^2$.
Consequently the vertex operators in the GSO($-$) sector, 
which represent the states of string stretching 
between the brane and the anti-brane, gets an additional piece 
that reflects the effect of the winding charge that the string
carries due to the stretching between the branes. If we denote by
$Y$ the world-sheet scalar along the direction of 
separation and if $\wt Y$
denotes the field dual to $Y$ then the additional piece in the
vertex operator is given by\footnote{One way to understand
eq.\refb{edelta} is to compactify the direction $y$ transverse
to the brane on a circle of large radius. Under T-duality this gets
mapped to a dual circle $\wt S^{1}$
of small radius, and the original 
D$p$-$\bd p$ system gets mapped to a D$(p+1)$-$\bd (p+1)$
brane configuration wrapped on the circle, with one of the branes
carrying a Wilson line proportional to $d'$ along $\wt S^1$. 
Thus an open string stretched between the brane and the
anti-brane will carry  momentum  proportional to $d'$ along
$\wt S^{1}$. With the normalization convention we have chosen
this momentum is equal to $\pm d'/2\pi$. Thus the vertex
operator representing these states 
will carry $e^{\pm i d' \wt Y}$ factors.}
\begin{equation} \label{edelta}
\Delta = e^{\pm i d' \wt Y/ 2\pi} = e^{\pm id \wt Y}
\end{equation}
where $d=d'/2\pi$ and
the $+$ and $-$ signs refer to sectors C and D respectively.

We shall now describe the off-shell 
vertex operators associated with the
string field components
we use in the level truncation analysis. 
First of all we have the
tachyon vertex operator.
For the non-BPS D-brane, the zero momentum 
tachyon vertex operator is given by 
\begin{equation}
\wh V_T = \xi c e^{-\phi} \otimes \sigma_1\, .
\end{equation} 
On a separated brane-antibrane pair the tachyon vertex operator
must carry the factors given in \refb{edelta}.
Since we require that the tachyon that 
condenses is real, the vertex operator must be hermitian. 
This gives
\begin{eqnarray} 
\wh V_T &=&  \xi c e^{-\phi} e^{id\wt Y}
\otimes \pmatrix{0&1 \cr 0&0} \otimes
 \sigma_1 + \xi c e^{-\phi} e^{-id\wt Y}
 \otimes \pmatrix{0&0 \cr 1&0} \otimes \sigma_1 \nonumber \\  
&=& \pmatrix{0& { \xi c e^{-\phi} e^{id
\wt Y}}\cr \xi c e^{-\phi} e^{-id\wt Y} &0} \otimes \sigma_1\, .
\end{eqnarray} 
$\wh V_T$ has total conformal weight $d^2-{1\over 2}$. Furthermore
we have
\begin{eqnarray}\label{Tbrst}
\wh Q_B  \wh V_T 
&=&  \left( {\left(d^2-\frac{1}{2}\right)\xi c 
\partial{c} e^{-\phi} 
e^{id\wt Y} - \frac{d}{\sqrt{2}}\psi_y c e^{id\wt Y}-
\eta e^{\phi}e^{id\wt Y}}\right)  
\otimes \pmatrix{0&1 \cr 0&0} \otimes i\sigma_2 \nonumber \\
&+& \left( {\left(d^2-\frac{1}{2}\right)\xi c \partial{c} 
e^{-\phi} e^{-id\wt Y} + \frac{d}{\sqrt{2}}\psi_y c 
e^{-id\wt Y}-\eta e^{\phi}e^{-id\wt Y}}\right)\otimes 
\pmatrix{0&0 \cr 1&0} \otimes i\sigma_2 \, , \nonumber \\
\end{eqnarray} 
\begin{eqnarray}\label{Teta}
\wh \eta_0 \wh V_T 
= \pmatrix {0& {c e^{-\phi}e^{id\wt Y}}\cr 
{c e^{-\phi}e^{-id\wt Y}} & 0} \otimes i\sigma_2 \, ,
\end{eqnarray} 
where $\psi_y$ is the world-sheet superpartner of $\wt Y$ on the
boundary.

The string field theory action has two $\ZZZ_2$ symmetries under
which the
tachyon vertex operator $\wh V_T$ is even. The first one corresponds
to
$Y\to -Y$, $\psi_y\to -\psi_y$ 
together with conjugation by the CP factor 
$\sigma_1\times I$. 
We shall denote the generator of this symmetry by $\sigma$.
The second one is the so called `twist symmetry' under which a
vertex operator with a conformal weight $h$ from the oscillators
(not counting the contribution from the $e^{\pm i d\wt Y}$ factors)
picks up a phase of $(-1)^{h+1}$ for integer $h$ and 
$(-1)^{h+{1\over 2}}$ for half integer $h$,
and the Chan-Paton factor associated
with this vertex operator gets transposed. We shall accompany
this by the $Y\to -Y$, $\psi_y\to -\psi_y$ 
transformation so that the tachyon vertex
operator $\wh V_T$ is even under this transformation. We shall
denote the generator of this symmetry by $\tau$.
We shall restrict to string field configurations
which are even under the $\sigma$ and $\tau$ transformations.

At the next level we have four more string fields
associated
with the vertex operators
\begin{eqnarray} \label{elist}
\wh V^{(1)}_K &=& \xi c e^{-\phi} \psi_y\otimes 
\pmatrix{1&0 \cr 0&-1} \otimes I \nonumber\\
\wh V^{(2)}_K &=&  \xi c e^{-\phi} \psi_y\otimes 
\pmatrix{1&0 \cr 0&1} \otimes I \nonumber\\
\wh V^{(1)}_M &=& c\partial{c} \xi \partial{\xi} 
e^{-2\phi}\otimes \pmatrix{1&0 \cr 0&-1}\otimes I \nonumber\\
\wh V^{(2)}_M &=&  c\partial{c} \xi \partial{\xi} 
e^{-2\phi}\otimes \pmatrix{1&0 \cr 0&1}\otimes I\, ,
\end{eqnarray} 
each of conformal weight 0.
Of these the vertex operators $\wh V^{(2)}_K$ and $\wh V^{(1)}_M$ are
odd under $\sigma$. Thus
we can set the components of the string field along this direction
to zero. The vertex operator $\wh V^{(2)}_M$ on the other hand is odd
under $\tau$. Thus we can set the coefficient of this operator also
to zero.
As a result we are left with only the vertex operator $\wh V^{(1)}_K$
which is even under both $\sigma$ and $\tau$.
The field associated with this vertex operator
has the interpretation of 
being the mode that shifts the branes in the opposite direction
by a distance proportional to its expectation value. 

{}From eq.\refb{elist} we get
\begin{eqnarray}\label{Kbrst}
\wh Q_B  \wh V^{(1)}_K 
&=&  \left(i\,\sqrt{2}\,c \partial{Y} 
+ \eta\psi_y e^{\phi} \right)  \otimes 
\pmatrix{1&0 \cr 0&-1} \otimes \sigma_3 \, ,
\end{eqnarray} 
\begin{eqnarray}\label{Keta}
\wh \eta_0  \wh V^{(1)}_K 
=  ce^{-\phi}\psi_y \otimes\pmatrix {1&0\cr0&-1} 
\otimes \sigma_3 \, .
\end{eqnarray}

\sectiono{Tachyon vacuum} \label{s3}

Now, with the ingredients prepared, we can
apply the method of level truncation 
to study tachyon condensation 
on separated branes. We shall use the 
expanded form (\ref{expanded}) of the Berkovits' action 
to evaluate the relevant terms at various levels. 
We define the level of a string field component 
multiplying a vertex operator of conformal 
weight $h$ to be $h + \frac{1}{2}$ so that 
the zero momentum tachyon at zero 
separation between the brane and the antibrane has weight zero.

\subsection{Level $d^2$ Computation}
The only field we need to keep in the analysis at the lowest 
level ($d^2$) is the tachyon field:
\be \label{ezero}
\wh \Phi = t \wh V_T\equiv \wh T\, .
\ee 
In order to get a non-vanishing correlation function
the total $\phi$ charge must add up to $-2$. 
This restricts the form of the pure tachyon 
potential to the form $at^2 +bt^4$, since terms involving
more than four powers of $\wh T$ (\i.e.\ $M+N>2$ 
terms in \refb{expanded}) vanish by $\phi$ charge conservation.

{}From the expanded form (\ref{expanded}), the quadratic term in 
the action reads
\begin{equation}
S_2 =  \frac{1}{8g^2}\langle\langle 
(\widehat Q_B \widehat T)( \widehat \eta_0 \widehat T ) \rangle\rangle
\end{equation}  
We use (\ref{Tbrst}) and (\ref{Teta}) to 
write down the form of the two-point function 
and then compute it using the standard correlation 
functions on the unit disk:\footnote{The magnitude 
of the correlation function is
easiest to compute on the disk; however to determine the sign we
map the disk to the upper half plane so that 
$f^{(n)}_i(0)<f^{(n)}_j(0)$
for $i<j$, and use the rules given in \refb{erule}.}
\ben%gin{eqnarray}
\langle\langle (\widehat Q_B \widehat T)( \widehat 
\eta_0 \widehat T ) \rangle\rangle &=& 2\left(d^2-\frac{1}{2}\right)
\bigg\{\langle\langle (\xi c \partial{c} e^{-\phi}e^{id\wt Y}) 
(c e^{-\phi}e^{-id\wt Y})\rangle\rangle \nonumber \\
&& \qquad \qquad \qquad  + 
\langle\langle (\xi c \partial{c} e^{-\phi}e^{-id\wt Y}) 
(c e^{-\phi}e^{id\wt Y})\rangle\rangle 
\bigg\} \, t^2 \, .\een
Using
\ben
\langle \langle(\xi c \partial{c} e^{-\phi}e^{\pm id\wt Y})
(c e^{-\phi}e^{\mp id\wt Y})\rangle\rangle &=&  
(f_1^{(2)\prime}(0)f_2^{(2)\prime}(0))^{d^2-\frac{1}{2}} 
\langle \xi c \partial{c} e^{-\phi}e^{\pm idX}(1)c 
e^{-\phi}e^{\mp idX}(-1)\rangle_{\hbox{disk}} \nonumber \\
&=&
-1  \een
we get
\ben
\langle\langle (\widehat Q_B \widehat T)
( \widehat \eta_0 \widehat T ) \rangle\rangle &=& -2(2d^2 - 1)t^2\, .
\end{eqnarray} 
This gives
\be \label{es2}
S_2 = -{1\over 2 g^2} \left( d^2 -{1\over 2}\right) t^2\, .
\ee

The quartic term in the action is
\begin{equation}
S_4 = - \frac{1}{48g^2} \left( \left\langle\left\langle 
(\widehat Q_B \widehat T)\widehat T( \widehat 
\eta_0 \widehat T )\widehat T \right\rangle\right\rangle - 
\left\langle\left\langle (\widehat Q_B \widehat T)
\widehat T \widehat T( \widehat \eta_0 \widehat T ) 
\right\rangle\right\rangle\right) \, .
\end{equation} 
Here we implicitly used the "twist symmetry" to simplify
the expression. 
We compute the correlation functions in the same way as above.
For example we have
\begin{eqnarray}
\langle\langle (\widehat Q_B \widehat T)\widehat T
( \widehat \eta_0 \widehat T )\widehat T \rangle\rangle = 
2\left\langle\left\langle (\eta e^{\phi} e^{id\wt Y}) (\xi c e^{-\phi} 
e^{-id\wt Y}) (ce^{-\phi} e^{id\wt Y}) ( \xi c e^{-\phi} e^{-id\wt Y})
 \right\rangle\right\rangle \nonumber  \\
+ 2\left\langle\left\langle (\eta e^{\phi} e^{-id\wt Y}) (
\xi c e^{-\phi}
 e^{id\wt Y}) (ce^{-\phi} e^{-id\wt Y}) (\xi c e^{-\phi} e^{id\wt Y}) 
 \right\rangle\right\rangle 
\end{eqnarray} 
and a similar expression for $\langle\langle 
(\widehat Q_B \widehat T)\widehat T\widehat T 
( \widehat \eta_0 \widehat T )\rangle\rangle$.
The calculation yields
\begin{equation}
g^2 S_4 =  -\frac{t^4}{12} \left(4+2\right) 
\end{equation}
The tachyon potential $V(t,d)$ is just the negative 
of the action as we have chosen the time 
coordinate to be periodic with a period 1. This gives
\begin{equation} \label{evtd}
V(t,d) = - (S_2+ S_4) = \frac{1}{2g^2}
\left[ \left( d^2-\frac{1}{2}\right)t^2 +
t^4 \right] 
\end{equation}  

We now minimize the potential with respect to t. This gives
\begin{equation}
\left[ (1-2d^2) - 4t^2\right] t = 0 
\end{equation} 
The solution $t_*$ to this equation corresponding to the
minimum of $V(t,d)$ is
\begin{eqnarray}
 t_*^2 &=& {1\over 4}\, (1-2d^2)
\quad \hbox{for $d^2 < 1/2$}  \\
t_* &=& 0  \quad \hbox{for $d^2 \geq 1/2$} .
\end{eqnarray} 
For $d^2 < 1/2$, the value of the potential at the minimum
is given by:
\begin{equation} \label{evmin}
V_{min} = -\frac{1}{32g^2} (1-2d^2)^2\, .
\end{equation}

For $d=0$ eq.\refb{evtd}-\refb{evmin} reduce to
\begin{eqnarray*}
V(t, d=0) = \frac{1}{g^2} \left( - \frac{1}{4}t^2 +
 \frac{1}{2} t^4\right) \\ 
t_*^2 = 1/4 \Rightarrow t_* = \pm 1/2 \\
V_{min} = -\frac{1}{32g^2}
\end{eqnarray*} 
These are in perfect agreement with the level zero 
calculations in \cite{0001084,0002211}.
%\end{itemize}

\subsection{Including the shift field}

Next we wish to compute the tachyon 
potential to the next non-trivial order. 
At this level we need to include the level ${1\over 2}$
shift field associated with the
vertex operator $\wh V_K^{(1)}$. Thus the string field has the expansion
\be \label{enextlevel}
\wh\Phi = t \wh V_T + \chi \wh V^{(1)}_K \equiv \wh T + \wh K\, .
\ee
We shall collect terms in the potential up to level $(1+2 d^2)$.
Again all terms with higher than four powers of the string field
vanish by $\phi$ charge conservation.
Thus we need to examine the terms of order 
$\chi^2$, $t^2\chi$ and $t^2 \chi^2$. Explicit computation 
to this order shows that the coefficient of the $\chi^2$ term
vanishes and the cubic and quartic terms are given by
\begin{equation}
S_3' = -{1\over \sqrt 2 \, g^2}\,
d \, \left(\frac{16}{27}\right)^{d^2-\frac{1}{2}} t^2 \chi \, ,
\end{equation}   
\begin{equation}
S_4' =  -\frac{1}{g^2} \left(2^{\frac{1}{2}-d^2} 
+ 2^{-1-2d^2}\right) t^2\chi^2\, .
\end{equation}
Adding $-S_3'-S_4'$ to the previous form 
of the tachyon potential we get
\begin{equation}
V(t,\chi,d) = \frac{1}{2g^2}\left[-
\left({\frac{1}{2}-d^2}\right)t^2 + t^4 
+ \sqrt{2}d\left(\frac{16}{27}\right)^{d^2-\frac{1}{2}} t^2\chi 
+ \left(2^{\frac{3}{2}-d^2}+2^{-2d^2}\right)
t^2\chi^2\right] 
\end{equation}
Eliminating $\chi$ using its equation of motion gives an
effective tachyon potential of the form:
\be \label{eeff}
V_{eff}(t) = \frac{1}{2g^2}\left[-
\left({\frac{1}{2}-d^2} + {1\over 2}
d^2 
\left(\frac{16}{27}\right)^{2d^2-1} \left(
2^{\frac{3}{2}-d^2}+2^{-2d^2}\right)^{-1}
\right)t^2 + t^4\right]\, .
\ee
{}From this we see that the effect of the field $\chi$ is
to make the tachyon more tachyonic. This is not surprising
since we expect that once a tachyon vacuum expectation value
is switched on, the field $\chi$ will develop a potential
that tries to pull the brane and the antibrane towards each
other. This is turn will reduce the tachyon mass$^2$.
Minimizing the effective potential \refb{eeff}
with respect to
$t$ now gives 
\begin{equation}
V_{min} = - \frac{1}{32g^2} \left[ 1 - 2 d^2 + d^2 
\left(\frac{16}{27}\right)^{2d^2-1} \left(
2^{\frac{3}{2}-d^2}+2^{-2d^2}\right)^{-1}
\right]^2 \, .
\end{equation} 

\begin{figure}
    \begin{center}
    \epsfysize 7cm
    \epsfbox{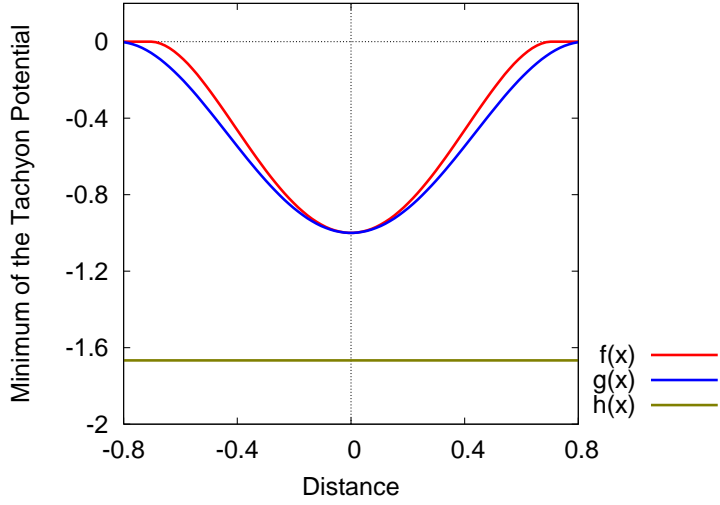}
      %\resizebox{120mm}{!}{\includegraphics{comparison.eps}}
      \caption{Plot of the minimum of tachyon potential 
      versus distance. We have plotted 
      $f(x)$= level $d^2$, $g(x)$= level 1/2  and
      $h(x)$= expected exact answer. We have used the unit $g^2=1/32$.}
      \label{comp2}
    \end{center}
  \end{figure}
On plotting the minimum of the potential as a function of $d$
(see Fig.~\ref{comp2}) we see that 
the dependence   on the distance $d$ is
less pronounced  after inclusion of the level 1/2 field
as compared to the case of the pure tachyon potential. 
We also see that the critical value $d_c$ of $d$ up to
which the tachyon potential has a minimum is now given by
the solution to the equation
\be\label{ecrit}
1 - 2 d_c^2 + d_c^2 
\left(\frac{16}{27}\right)^{2d_c^2-1} \left(
2^{\frac{3}{2}-d_c^2}+2^{-2d_c^2}\right)^{-1}=0\, .
\ee
This is larger than the original value $1/\sqrt 2$ indicating that
the after taking into account corrections due to higher level fields the
minimum of the potential remains below zero for a 
larger range of $d$.\footnote{We should note however that for
$d^2\ge 1/2$ the level $(1+2d^2)$ terms in the potential are of
higher level than the level 2 terms arising from the $\chi^4$ terms.
Thus for a consistent approximation we must also include the
quartic coupling of the shift field.}
We expect the dependence  on $d$ to reduce as we 
include more and more fields into the analysis. 
The true minimum of the tachyon potential should not 
depend on what distance we separate out the branes initially. 
This has been shown schematically in Fig.~\ref{comp2} .

%\end{itemize}

\sectiono{Rolling non-universal tachyon} \label{s4}

In this section we shall set up the formalism for describing the
rolling tachyon solution on a separated brane-antibrane pair.
We begin by
considering an unstable D-brane system in bosonic string theory,
containing a primary boundary operator $V$ of dimension $h<1$. This
would correspond to a tachyon of mass$^2=(h-1)$ in the spectrum.
Our goal will be to construct a time dependent solution in the
theory that describes the rolling of this tachyon away from the
maximum of its potential. 

Let $X=-i X^0$ denote the world-sheet scalar
associated with the Wick rotated
time coordinate. Then the operator
\be \label{e1}
\wt V = V \, e^{i\sqrt{1-h} X}\, ,
\ee
is an operator of dimension 1. This will be an exactly marginal operator
if a product of arbitrary number of these operators does not contain an
operator of dimension 1. Since a product of $n$ of these operators
contains a factor of $e^{in\sqrt{1-h} X}$ of dimension $n^2(1-h)$,
we see that the condition for exact marginality is easily achieved
if $n^2(1-h)>1$ for $n\ge 2$, \i.e. if 
\be \label{e2}
h<{3\over 4}\, .
\ee 
Since 
\be \label{e2a}
\wt V(u) \wt V(u') = K (u-u')^{2 - 4h} e^{2i\sqrt{1-h} X(u')} 
+ \cdots
\ee
with $K$ denoting a constant and
$\cdots$ denoting less singular terms, we see that for $h$ satisfying
\refb{e2} the power
of $(u-u')$ in $V(u) V(u')$\
is larger then $-1$. 

Consider now deforming the theory by adding to the action the term
\be \label{e3}
\lambda\, \int du \, \wt V(u)\, .
\ee
Then the correlation functions in the deformed theory are computed
by inserting into the correlation function of the undeformed theory
the operator
\be \label{e4}
\exp\left[ - \lambda\, \int du\, \wt V(u) \right]\, .
\ee
After expanding the exponential factor we encounter multiple
integrals of the form
\be \label{e5}
\int du_1\int du_2\cdots \int d u_n \, \wt V(u_1)\cdots \wt V(u_n)\, .
\ee
Since the operator product $\wt V(u_i) \wt V(u_j)$ is less singular
than $(u_i - u_j)^{-1}$, the above integral, inserted into a correlation
function of the undeformed theory, gives completely regular 
integrals.\footnote{The exceptions are correlation functions of
boundary operators whose product with $\wt V(u)$ have
stronger than $(u-u')^{-1}$ singularity. 
Such boundary operators must be
renormalized in the deformed theory although the deformed
theory itself
is finite. In our analysis we shall consider correlation
functions of bulk operator(s) in the deformed theory, inserted at
point(s) away from the boundary. Hence
we do not encounter the problem mentioned above.}
Thus the correlation functions of the deformed theory are determined
in terms of the correlation functions of the undeformed theory without
any need to regularize the theory.

The operator $\wt V$ defined in \refb{e1} is of course not 
hermitian and hence the deformation \refb{e4} does not
produce a physical background of the open string theory.
This problem disappears after inverse Wick rotation 
$X\to -i X^0$. In this case the operator $\wt V(u)$ becomes
$V(u) \, e^{X^0(u)\sqrt{1-h^2}}$ and the deformed theory
describes a physical open 
string background. This in fact describes the rolling of the
tachyon associated with the operator $V$ away from its 
maximum.

Generalization to the case of unstable D-branes in
superstring theory is straightforward.
Suppose we have a vertex operator described by the
superfield $V_{-1}+\theta V_0$ 
with $(V_{-1}, V_0)$ having dimensions
$(h, h+ {1\over 2})$ and $\theta$ denoting the fermionic
world-sheet coordinate. If $V_{-1}$ is a superconformal
primary then this describes a tachyon on the D-brane world-volume
of mass$^2=(h-{1\over 2})$. 
We now denote as before by $X=-i X^0$ the 
world-sheet scalar
associated with the Wick rotated
time coordinate, and by $\psi=-i \psi^0$ its fermionic
superpartner.  Then the vertex operator
\be \label{e6a}
\wt V(\theta) = 
e^{i\, (X+\theta \psi)\, \sqrt{{1\over 2} -h} } 
\left( V_{-1} + \theta V_0\right) \equiv (\wt V_{-1} + \theta \wt V_0)
\, 
\ee
describes a primary
superfield whose lowest component has dimension 1/2. Here
\be \label{e6}
\wt V_{-1} = \wt V(\theta)|_{\theta=0} 
= e^{i  X\sqrt{{1\over 2} - h}} \, V_{-1}, \qquad
\wt V_0 = \int d\, \theta \, \wt V(\theta)
= e^{i  X\sqrt{{1\over 2} - h}} \left(V_0 + i  \psi V_{-1}
\sqrt{{1\over 2} - h}
\right)
\, ,
\ee
Hence the highest component $\wt V_0$ of the superfield
$\wt V(\theta)$
is
marginal. In order for it to be exactly marginal we need its
operator product with itself not to contain any other
marginal deformation. Repeating the analysis for
bosonic string theory we see that this can be guaranteed
if
\be \label{e7}
h < {1\over 4}\, .
\ee
Furthermore in this case the operator product of \refb{e6}
with itself has a singularity that is milder than
$(u-u')^{-1}$. Thus the correlation functions in the
theory deformed by the operator
\be \label{e8}
\lambda\, \int du \, \wt V_0(u)\, ,
\ee
are unambiguously determined in terms of the correlation
functions in the undeformed theory without any need to regularize
the theory. 

As in the case of bosonic string theory,  the
operator $\wt V_0$ defined in \refb{e6} is not hermitian
and hence deformation \refb{e8} does not represent a physical
open string background. However by the inverse Wick
rotation $iX\to X^0$, $i\psi \to \psi^0$ we get a
hermitian vertex operator
\be \label{e8a}
\wt V_0 
= e^{ X^0\sqrt{{1\over 2} - h}} \left(V_0 + \psi^0 V_{-1}
\sqrt{{1\over 2} - h}
\right)
\, .
\ee
Thus the deformation of the original theory by this operator
produces a physical open string background.

We conclude this section with two examples. The first example
is a D$p$-$\bd$$p$ brane wrapped on a circle of radius $R$.
Let $y$ denote the coordinate along the compact circle and $Y$
and $\psi_y$ be the associated world-sheet scalar and fermion
fields respectively.
For $R>\sqrt 2$ there is a tachyon of 
mass$^2={1\over R^2} -{1\over 2}$
described by the vertex operator:
\be \label{e9}
V_{-1}+\theta V_0 = \sigma_1\,
\cos\left({Y+ \theta \psi_y \over R}\right) = \sigma_1\,
\left(\cos{Y\over R} -
{1\over R}\, \theta \,\psi_y \,\sin{Y\over R}\right)
 \, ,
\ee
where the Pauli matrix $\sigma_1$ represents the external 
Chan-Paton factor.\footnote{Note that we have
dropped the `hat' and the 
internal CP factor from the vertex operator. It plays no role
in our analysis since we always have even number of GSO($-$)
operators in a correlator and hence the trace over the product of
internal CP factors will
always give an overall factor of 2. This can be absorbed into
the normalization of the disk partition function.}
$V_{-1}$ has dimension $h=R^{-2}$.
Thus \refb{e6},
\refb{e7} shows that for 
\be \label{e10}
R>2\, 
\ee
we can construct an exactly marginal deformation generated
by the
operator\footnote{For $R=\sqrt 2$ the effect of switching on this
operator was analyzed in \cite{0207105,0212248}. 
Although it appears to be the
sum of two different exactly marginal operators each of which
gives a solvable deformation, these operators anticommute and hence
the resulting theory does not appear to be solvable via known
methods.
}
\be \label{e11}
\wt V_0 
= \sigma_1\, 
e^{i  X\sqrt{{1\over 2} - {1\over R^2}}} \,
\left(-\psi_y\, {1\over R}\, \sin{Y\over R}+  i \psi
\sqrt{{1\over 2} - {1\over R^2}} \, \cos{Y\over R}
\right)\, .
\ee
The operator product  $\wt V(u) \wt V(u')$ is less singular
than $(u-u')^{-1}$ and hence the correlation functions of
the deformed theory are free from any singularity.

The second example, which we shall analyze in detail in
later sections,
is a D$p$-$\bd p$ brane pair separated by a distance $2\pi d$
in the transverse direction. 
Let $y$ denote the
transverse coordinate along the direction of separation
of the branes, $\wt Y$ denote the world-sheet scalar dual 
to the scalar field $Y$ associated with
the coordinate $y$, and $\wt\psi_y$ denote the
fermionic superpartner of $\wt Y$.
In this case for $d<{1\over \sqrt 2}$
there is a tachyonic mode on this system, represented by the
vertex operator
\be \label{e12}
V_{-1}+\theta V_0 = \sigma^+ \,
e^{i (\wt Y+ \theta \wt\psi_y)\, d} 
+ \sigma^- \,
e^{-i (\wt Y+ \theta \wt\psi_y)\, d} 
= \sigma^+ \,
(1 + i\theta \wt\psi_y\, d)\, e^{i \wt Y\, d} 
+ \sigma^- \,
(1 - i\theta \wt\psi_y \, d)\, e^{-i \wt Y\, d}\, ,
\ee
where  $\sigma^\pm$ are the external
Chan-Paton factors
\be \label{e13}
\sigma^+ = \pmatrix{0 & 1\cr 0 & 0}, \qquad
\sigma^- = \pmatrix{0 & 0\cr 1 & 0}\, .
\ee
Since $V_{-1}$ has dimension $h=d^{2}$ we see from
\refb{e6}, \refb{e7} that for 
\be \label{e14}
d<{1\over 2}\, 
\ee
we can construct an exactly marginal deformation generated
by the
operator
\be \label{e15}
\wt V_0 
= i\,\sigma^+\, 
e^{i  X\sqrt{{1\over 2} - { d^2}}} \,
e^{i \wt Y\, d}\, \psi^+ + i\,\sigma^-\, 
e^{i  X\sqrt{{1\over 2} - {d^2}}} \,
e^{-i \wt Y\, d}\, \psi^-
\ee
where
\be \label{e15a}
\psi^+ = \left(\wt\psi_y\, d+   \psi
\sqrt{{1\over 2} - {d^2}}
\right), \qquad 
\psi^- = \left(- \wt\psi_y\, d+   \psi
\sqrt{{1\over 2} - {d^2}}
\right)\, .
\ee
Again the operator product  $\wt V(u) \wt V(u')$ is less singular
than $(u-u')^{-1}$ and hence the correlation functions of
the deformed theory are free from any singularity.

\sectiono{Time dependence of pressure on a 
separated brane-antibrane
system with a rolling tachyon} \label{s5}

We consider a brane-antibrane system separated by a distance
$2\pi d$ with
$d<{1\over 2}$ in the presence of a rolling tachyon 
background generated by the deformation 
\be \label{e16}
\lambda \int du \, \wt V_0(u)\, ,
\ee
with $\wt V_0$ given in \refb{e15}.  Let $p(x)$ denotes the
$x$-dependent tangential
pressure of the brane
generated by this deformation and
$p_0$ be the pressure in the absence of this deformation.
$p(x)/p_0$ has a Fourier expansion of the form
\be \label{e17}
p(x)/ p_0 = \sum_{n\ge 0} a_n 
e^{i n x \sqrt{{1\over 2} - {d^2}}}\, ,
\ee
for constants $a_n$. The coefficients $a_n$ can be found by
examining the boundary state of the deformed brane if it is
known, or
equivalently from the disk one point function of the matter
vertex operator
$e^{-i n X\sqrt{{1\over 2} - {d^2}}}$
inserted at the center of the unit disk in the deformed 
theory:\footnote{The full boundary state contains matter
and ghost parts, but the ghost part of the correlation function 
as well
as the matter part involving fields other than $X$, $\wt Y$
and their fermionic partners cancel
between $p(x)$ and $p_0$, leaving behind only the part
involving $X$, $\wt Y$
and their fermionic partners.}
\be \label{e19}
a_n = \langle e^{-i n X(0)\sqrt{{1\over 2} - {d^2}}} 
\rangle_{deformed}\, ,
\ee
where  $\langle~\rangle_{deformed}$ denotes the correlation
function in the deformed theory, normalized such that in the
undeformed theory the disk partition function is 1. 
Representing the deformation of the Euclidean world-sheet
action as
\be \label{e20}
\lambda \, \int du\wt V_0(u)\, ,
\ee
with $u$ labeling the coordinates on the boundary of the
disk, we can reexpress $a_n$ as\cite{0212248}
\be \label{e21}
a_n
=
{1\over 2}
 \sum_{r=0}^\infty {\lambda^{2r}\over (2r)!} 
 \, \int du_1 \cdots d u_{2r}\, 
 \left\langle e^{-i n X(0)\sqrt{{1\over 2} - {d^2}}} 
 Tr\left(\wt V_0(u_1) \cdots \wt V_0(u_{2r})\right)
\right\rangle_0\, ,
\ee
where $\langle ~\rangle_0$ denotes the correlation function
in the undeformed theory on the unit disk
and $Tr$ denotes trace over the
Chan-Paton factors. In \refb{e21} we have used the fact that
the trace over the Chan-Paton factor vanishes if we have
odd number of $\wt V_0$ insertions on the boundary. 
The overall factor of $1/2$ is a reflection of the factor of 2
appearing in the expression for the unperturbed pressure $p_0$
from the trace over the Chan-Paton factors.
Using
the results
\be \label{e22}
\left(\sigma^+\right)^2 = \left(\sigma^-\right)^2 = 0\, ,
\ee
we see that only two strings of Chan-Paton factors
contribute to the correlation function -- $Tr(\sigma^+
\sigma^-\sigma^+\sigma^-\cdots)$ and 
$Tr(\sigma^-
\sigma^+\sigma^-\sigma^+\cdots)$. The associated vertex
operators must be cyclically ordered on the boundary of the
unit disk. 
Both strings give the same contribution.
Finally $X$-momentum conservation, together
with the fact that each of the $\wt V_0$ carries $X$-momentum
$\sqrt{{1\over 2} - {d^2}}$, shows that the correlator
\refb{e21} is non-vanishing only when $n=2r$.
Using this we can express \refb{e21} as 
\ben \label{e23}
a_{2r+1} &=& 0 \quad \hbox{for $r\in\ZZZ$}\nonumber \\
a_{2r} &=& \lambda^{2r}\,
\int_0^{2\pi} \, dt_1 \int_0^{t_1} \,
dt_2\cdots
\int_0^{t_{2r-1}} \, dt_{2r} \nonumber \\
&& \Bigg\langle 
e^{-2 i r X(0)\sqrt{{1\over 2} - {d^2}}} \,
e^{i X(u_1) \sqrt{{1\over 2} - {d^2}}}
\cdots e^{i X(u_{2r}) \sqrt{{1\over 2} - {d^2}}}
\nonumber \\ && \times 
\, e^{i \wt Y(u_1)\, d} \, e^{-i \wt Y(u_2)\, d}
\cdots e^{i \wt Y(u_{2r-1})\, d} e^{-i \wt Y(u_{2r})\, d}
\nonumber \\
&& \times\, 
(i)^{2r} \, \psi^+(u_1) \psi^-(u_2) \psi^+(u_3)
\psi^-(u_4) \cdots \psi^+(u_{2r-1})\psi^-(u_{2r})
\Bigg\rangle\,, \qquad u_i\equiv e^{it_i}\, , \nonumber \\
&& \quad \hbox{for $r\in\ZZZ$, $r\ge 0$}\, . 
\een
Note that we have replaced $\int \prod d u_i$ by 
$\int\prod_i dt_i$, -- this
requires that in computing the correlators on the right hand side
of \refb{e23} all the fields need to be defined in the $t$-coordinate
system.
The part of the correlator involving the scalar fields 
$X$ and $\wt Y$
gives
\be \label{e24}
\prod_{i<j} |u_i - u_j|^{1 - 2{d^2}
+ 2(-1)^{i+j} {d^2}} = \prod_{i<j} 
\left|2\sin{t_i -t_j\over 2}\right|^{1 - 2{d^2}
+ 2(-1)^{i+j} {d^2}}\, .
\ee
On the other hand the fermionic correlators can be calculated
with the help of Wick's theorem using the two point 
functions\footnote{The extra factor of $i e^{i(t_i + t_j)/2}$ comes
from the conformal transformation of $\psi^\pm$ from the 
$u$ to $t$ coordinate.}
\ben \label{e25}
&& \langle \psi^+ (u_1) \psi^+(u_2)\rangle 
= \langle\psi^- (u_1) \psi^-(u_2) \rangle
= {i\over 2}\,
{e^{i(t_i+t_j)/2} \over u_1 - u_2} = {1\over 2} {1
\over 2\sin{t_i -t_j\over 2}}, \nonumber \\
&& 
\langle\psi^+ (u_1) \psi^-(u_2)\rangle
= i\, \left( {1\over 2} - 2 \, d^2\right) \, {e^{i(t_i+t_j)/2}
\over u_1 - u_2} =\left( {1\over 2} - 2 \, d^2\right)
\, {1
\over 2\sin{t_i -t_j\over 2}}\, .
\een
Although the fermionic correlator in eq.\refb{e23} contains
many terms we can organize them in a compact form by
collecting all terms in which a certain number (say $s$) of
$\psi^+$ pairs have been contracted with each other, an equal
number of $\psi^-$ pairs have been contracted with each other,
and the $(r-2s)$ left-over $\psi^+$'s have been contracted
with $(r-2s)$ left-over $\psi^-$'s. 
After taking into account the combinatoric factors 
we can express \refb{e23} as
\ben \label{e27}
a_{2r} &=& \, (-1)^{r}\, \lambda^{2r}\,
\int_0^{2\pi} \, dt_1 \int_0^{t_1}\,
dt_2\cdots
\int_0^{t_{2r-1}} \, dt_{2r}  
\, \prod_{i<j} \left|2 \sin {t_i - t_j\over 2} \right| ^{1 - 2{d^2}
+ 2(-1)^{i+j} {d^2}}\nonumber \\  
&&
 \sum_{s=0}^{[r/2]} {(1 - 4 d^2)^{r - 2s} (-1)^s
\over (r-2s)! (s!)^2 2^{2s}2^r}\,
\Bigg[\prod_{l=2s}^{r-1}
\left( 2 \sin {t_{2l+1} - t_{2l+2}\over 2}\right)^{-1}
\nonumber \\
&& \prod_{k=0}^{s-1}  
\left\{\left(2 
\sin {t_{4k+1} - t_{4k+3}\over 2} \right )^{-1}
\left(2 
\sin {t_{4k+2} - t_{4k+4}\over 2} \right )^{-1}\right\}
\nonumber \\
&& + (-1)^{P+P'} \times \hbox{permutations $P$ of
$t_1,t_3,\cdots t_{2r-1}$} 
\times \hbox{permutations $P'$ of
$t_2,t_4,\cdots t_{2r}$}\Bigg]\, , \nonumber \\
\een
where $[r/2]$ denotes the integral part of $r/2$.

Another compact representation of the correlators is provided
by the superfield representation\cite{0212248}. If we denote by 
$\wt V_\pm(u,\theta)$
the superfields
\be \label{e27a}
V_{\pm}(u,\theta) = e^{i (X + \theta\psi) \sqrt{{1\over 2} - d^2}}\, 
e^{\pm i (\wt Y+ \theta \wt\psi_y)\, d}\, ,
\ee
then the expression for $a_{2r}$ given in \refb{e23} can be 
written as
\ben \label{e27b}
a_{2r} &=& \lambda^{2r}\,
\int_0^{2\pi} \, dt_1 \int_0^{t_1} \,
dt_2\cdots
\int_0^{t_{2r-1}} \, dt_{2r} \int d\theta_1\cdots \int d\theta_{2r}
\nonumber \\
&& \left\langle e^{-2i r X(0)}\,
\wt V^+(u_1, \theta_1) \wt V^-(u_2, \theta_2) 
\wt V^+(u_3, \theta_3) \wt V^-(u_4, \theta_4) 
\cdots
\wt V^+(u_{2r-1}, \theta_{2r-1}) \wt V^-(u_{2r}, \theta_{2r}) 
\right\rangle \nonumber \\
&=& \lambda^{2r}\,
\int_0^{2\pi} \, dt_1 \int_0^{t_1} \,
dt_2\cdots
\int_0^{t_{2r-1}} \, dt_{2r} \int d\theta_1\cdots \int d\theta_{2r}
\nonumber \\ &&
\qquad \qquad \prod_{i<j} \left[ 2 \sin{t_i - t_j\over 2}+ \theta_i \theta_j
\right]^{1 - 2{d^2}
+ 2(-1)^{i+j} {d^2}}
\, . %\nonumber \\
\een
After integration over $\theta_i$ eq.\refb{e27b} reduces to
\refb{e27}.

Once the coefficients $a_n$ have been calculated, we can
use \refb{e17} to calculate $p(x)$. In fact after the
inverse Wick rotation $ix\to x^0$ we get
\be \label{e28}
p(x) / p_0 =   \sum_{r\ge 0} a_{2r} e^{2r x^0
\sqrt{{1\over 2} - d^2}}\, .
\ee
Since $a_{2r}\propto \lambda^{2r}$, we can absorb $\lambda$
into an additive constant in $x^0$. Thus by suitable choice of the
origin of the $x^0$ coordinate we can set 
$\lambda=(\sqrt 2\pi)^{-1}$. 
In this
case \refb{e27} reduces to 
\ben \label{e29}
a_{2r} &=& \, (-1)^{r}\, 
\int_0^{2\pi} \, {dt_1\over 2\pi} \int_0^{t_1}\,
{dt_2\over 2\pi} \cdots
\int_0^{t_{2r-1}} \, {dt_{2r}\over 2\pi}  
\, \prod_{i<j} \left|2 
\sin {t_i - t_j\over 2} \right| ^{1 - 2{d^2}
+ 2(-1)^{i+j} {d^2}} \nonumber \\
&&\sum_{s=0}^{[r/2]} {(1 - 4 d^2)^{r - 2s} \, (-1)^s\, 
\over (r-2s)! (s!)^2 2^{2s}}\,
 \Bigg[\prod_{l=2s}^{r-1}
\left( 2 \sin {t_{2l+1} - t_{2l+2}\over 2}\right)^{-1}
\nonumber \\
&& \prod_{k=0}^{s-1}  
\left\{\left(2 
\sin {t_{4k+1} - t_{4k+3}\over 2} \right )^{-1}
\left(2 
\sin {t_{4k+2} - t_{4k+4}\over 2} \right )^{-1}\right\}
\nonumber \\
&& + (-1)^{P+P'} \times \hbox{permutations $P$ of
$t_1,t_3,\cdots t_{2r-1}$} 
\times \hbox{permutations $P'$ of
$t_2,t_4,\cdots t_{2r}$}\Bigg]\, . \nonumber \\
\een

\sectiono{Numerical results} \label{s6}

\begin{figure}
    \begin{center}
    \epsfysize 7cm
    \epsfbox{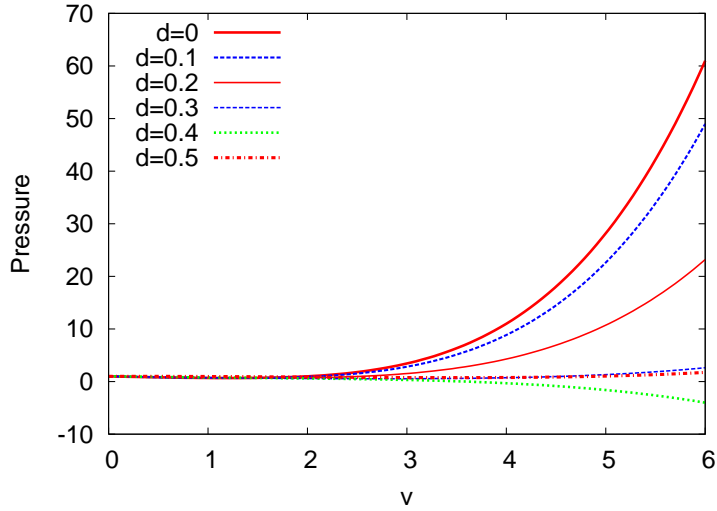}
      %\resizebox{120mm}{!}{\includegraphics{comparison.eps}}
      \caption{Plot of the pressure as a function of $v=
      e^{2\sqrt{{1\over 2} - d^2} x^0}$ for various values of $d$.}
      \label{fig5}
    \end{center}
  \end{figure}

We have evaluated the first few coefficients given in \refb{e29}
using Monte Carlo integration techniques. The results are given in
table \ref{t1}. The integrand is generated 
using a code in {\it {Mathematica}}
and then we use VEGAS\cite{vegas} to do the multidimensional integrals. 
Since these give the first few terms in the expansion
of $p(x^0)/p_0$ in a power 
series in $v\equiv e^{2x^0\sqrt{{1\over 2}-d^2}}$, we cannot reliably
estimate the late time behaviour of $p(x^0)$ using these results.
A plot of pressure as a function of 
$v\equiv e^{2x^0\sqrt{{1\over 2}-d^2}}$ is shown in 
Fig.~\ref{fig5}. {}From this it seems that the function does not display
the wild oscillation of the kind seen in the tachyon profile computation
associated with the rolling tachyon solution in 
string field theory; instead
it may have a finite radius of convergence, and may admit an analytic
continuation to infinite time as in the case of the behaviour of the
pressure in the $d=0$ case. With our present data it is not possible to
make any reliable estimate of the asymptotic value of the pressure.
Nevertheless we give in table \ref{t2}
the results of fitting a ratio of quadratic
functions of $v$ to the series expansion. For the $d=0$ case the
exact answer is known and it agrees with the result given in the
table. 
If we take these results seriously
then within numerical errors
these results are consistent with the hypothesis that at late time the
pressure vanishes.

\begin{table} {\small
\begin{tabular}{|c||c|c|c|c|c|}
\hline  & r=0 & r=1 & r=2 & r=3 & r=4 \\
\hline d=0 & 1 & 0.4999999 (1.5E-07) 
& 0.2500775 (3.5E-05) & 0.1249626 (2.3E-05) & 0.06246 (1.0E-04) \\ 
\hline d=0.1 & 1 & 0.4803255 (1.6E-07) 
& 0.2284956 (3.2E-05) & 0.1079812 (2.0E-05) & 0.05088 (1.8E-05) \\ 
\hline d=0.2 & 1 & 0.4250437 (1.8E-07) & 0.1719809 
(2.4E-05) & 0.0671200 (1.4E-05) & 0.02550 (1.3E-05) \\ 
\hline d=0.3 & 1 & 0.3442298 (3.5E-07) & 
0.1003526 (1.8E-05) & 0.0243640 (9.6E-05) & 0.00413 (2.2E-05) \\ 
\hline d=0.4 & 1 & 0.2512369 (1.0E-05) & 
0.0338092 (5.0E-05) & -0.0039147 (2.9E-05) & -0.00429 (2.8E-05) \\ 
\hline d=0.5 & 1 & 0 & -0.0416668 (1.5E-07) & 
0 & 0.001729 (1.2 E-05) \\ 
\hline  
\end{tabular} }
\caption{The table containing the coefficients 
$(-1)^ra_{2r}$ calculated
from \refb{e29}. 
The numbers in the parenthesis are the estimated 
errors of the numerical calculations. \label{t1}}
\end{table}

\begin{table} {\small
\begin{tabular}{|c||c|c|}
\hline  Separation & (2,2) Pade Approximation & Late Time Pressure  \\ 
\hline d=0 & $ \{1 + 2 k v\}/\{1 + (1/2 + 2k) v + k v^2\}$ & 0 \\ 
\hline d=0.1 & $ \{1 + (0.20032) v + (0.00052) v^2\}/
\{1 + (0.68065) v + (0.09895)v^2\}$ & 0.00526 \\ 
\hline d=0.2 & $ \{1 + (0.23555) v + (0.00067) v^2\}/\{1 
+ (0.6606) v + (0.10945)v^2\}$ & 0.00612 \\ 
\hline d=0.3 & $ \{1 + (0.27651) v - (0.00325) v^2\}/\{1 
+ (0.6207) v + (0.11007)v^2\}$ & -0.02936 \\ 
\hline d=0.4 & $ \{1 + (0.19569) v - (0.00283) v^2\}/\{1 
+ (0.4469) v + (0.07565)v^2\}$ & -0.03741 \\ 
\hline d=0.5 & $ \{1 - (0.000177) v^2\}/\{1 + (0.04149)v^2\} $ 
& -0.00427 \\ 
\hline 
\end{tabular} }
\caption{The table containing the Pade approximant results
for the function representing $p(x^0)/p_0$.  
Here $v\equiv e^{2x^0\sqrt{{1\over 2}-d^2}}$.
The last column gives
the late time behaviour of the pressure if we take these expressions
seriously. In the first row $k$ is
an arbitrary constant. \label{t2}}
\end{table}

\sectiono{Discussion} \label{s7}

We have seen in \S\ref{s5} that given a tachyon with mass$^2$
less than a certain critical value we can generate a deformation of the
original CFT by an exactly marginal operator describing a rolling
tachyon solution. Furthermore the deformed theory does not require
any additional renormalization beyond those required to renormalize
the original CFT.

It turns out that precisely for these ranges of tachyon mass$^2$
we can generate a rolling tachyon solution of open string field
theory following the method of 
\cite{0701248,0701249,0704.0930,0704.0936, 0704.2222,0704.3612, 
0705.0013,0706.0717,0710.1342,0801.0573}.\footnote{A
general method for constructing a solution of open bosonic string
theory describing arbitrary marginal deformation has been developed
in \cite{0707.4472,0708.3394}, but this requires adding
`counterterms' and the procedure is more complicated.}
Let us first consider the case of open bosonic string theory. In this
case if we have a matter sector vertex operator $\wt V$ of dimension
1, then we can generate a non-singular solution of open bosonic
string field theory provided integrals
of the form $\int du' \wt V(u) \wt V(u')$ 
do not diverge in the region $u'\simeq u$\cite{0701248,0701249}. 
But this is precisely the condition that
$\wt V(u) \wt V(u')$ will have a singularity softer than
$(u-u')^{-1}$. Similarly the condition under which one can generate
a non-singular solution in open superstring field theory corresponding
to a dimension half matter vertex operator $\wt V_{-1}$ and its
dimension 1 superpartner $\wt V_0$ is that
$\int du' \wt V_{-1}(u) \wt V_0(u')$ and 
$\int du' \wt V_{0}(u) \wt V_0(u')$ do not diverge from $u'\simeq
u$ region\cite{0704.0930,0704.0936,0704.3612}.
For $\wt V_{-1}$ and $\wt V_0$ given in eq.\refb{e6} this
happens precisely for $h<{1\over 4}$, \i.e.\ when \refb{e7} is
satisfied.  Thus both for bosonic and superstring field theory we
can construct a non-singular solution describing a rolling 
non-universal tachyon when the  boundary CFT associated
with the solution can be defined without any need to regularize
and renormalize the theory.
It may be of interest to study these solutions numerically.

\bigskip

{\bf Acknowledgements:} We would like to thank 
T.~Erler for useful discussions and Girish Kulkarni 
and especially V. Ravindran for help in the numerical part of the work.

%\small
%\baselineskip 13.7pt

\end{document}